\documentclass[a4paper, usenatbib]{mnras}
\usepackage{graphicx}
\usepackage{epsfig}
\usepackage{subfig}
\usepackage{multirow}
\usepackage{mathrsfs,amsmath} 
\usepackage[utf8]{inputenc}

\title[Calibration Requirements for EoR Observations]{Calibration Requirements for Epoch of Reionization 21-cm signal observations - I. Effect of time-correlated gains}

\author[J. Kumar et al.]{Jais Kumar$^{1}$\thanks{E-mail: jaisk.rs.phy16@iitbhu.ac.in},
Prasun Dutta$^{1}$ and
Nirupam Roy$^{2}$
\\
$^{1}$Department of Physics, Indian Institute of Technology (Banaras Hindu University), Varanasi - 221005, India\\
$^{2}$Department of Physics, Indian Institute of Science, Bangalore, 560012, India
}
\date{Accepted XXX. Received YYY; in original form ZZZ}
\pubyear{2020}

\begin{document}
\label{firstpage}
\pagerange{\pageref{firstpage}--\pageref{lastpage}}
\maketitle
\begin{abstract}
The residual gain errors add to the systematics of the radio interferometric observations. In case of the high dynamic range observations, these systematic effects dominates over the thermal noise of the observation. In this work, we investigate the effect of time-correlated residual gain errors in the estimation of the power spectrum of the sky brightness distribution in high dynamic range observations. Particularly, we discuss a methodology to estimate the bias in the power spectrum estimator of the redshifted 21-cm signal from neutral hydrogen in the presence of bright extragalactic compact sources. We find, that for the visibility based power spectrum estimators, particularly  those use nearby baseline correlations to avoid noise bias, the bias in the power spectrum arises mainly from the time correlation in the residual gain error. The bias also depends on the baseline distribution for a particular observation. Analytical calculations show that the bias is dominant for certain types of baseline pairs used for the visibility correlation. We perform simulated observation of extragalactic compact sources in the presence of residual gain errors with the GMRT like array and estimate the bias in the power spectrum. Our results indicate that in order to estimate the redshifted 21-cm power spectrum, better calibration techniques and estimator development are required.
\end{abstract}

\begin{keywords}
cosmology: dark ages, reionization -- methods: analytical, numerical, statistical -- techniques: interferometric
\end{keywords}

\section{Introduction}
Interferometers are the key instruments to measure multi scale properties of the sky brightness distribution at radio frequencies. An interferometer consists of one \citep{Ramesh} or many receiving elements in the form of antenna (GMRT\footnote{The Giant Metrewave Radio Telescope} \citep{GMRT}, VLA \footnote{The Karl G. Jansky Very Large Array} \citep{VLA} etc.) or dipoles (the LOFAR \footnote{The LOw Frequency ARray} \citep{LOFAR}, the MWA \footnote{The Murchison Widefield Array} \citep{MWA}, etc.). Each of these elements measures one or two components of the electric field of the incoming electromagnetic waves. The electric fields from different elements are then correlated and averaged over certain time intervals. The later, usually known as the visibility, is the spatial correlation function of the electric field at the earth's surface \citep{Thompson&Moran}. The electric field measured by each element of the interferometer is also affected by the ionospheric electron density variation, the receiver beam and the electronics following the receiver. Collectively, we mention these effects as antenna based gains. The gain usually varies with time and frequency \citep{SynthesisImaging}. Several procedures are used during observation and data reduction to estimate the gains and calibrate the visibilities. These include observation of the standard calibrator sources to measure long time scale variation of the gain, self-calibration \citep{Selfcal} to reduce the effect of rapid ionospheric changes, redundancy calibration \citep{Noordam, Weiringa}, etc. For the telescopes with large field of view, additionally, direction-dependent calibration techniques are used to estimate and mitigate the effect of ionosphere along the different directions over each antenna \citep{van-der-Tol, Wijnholds}.

Different aspects of the radio interferometric observations, like the rapid fluctuation in gain with time, measurement noise in each visibilities, observation overhead on the calibrators, etc, limits the accuracy of the calibration in every interferometric observation. However, with modern algorithms the calibration errors are rather small and high dynamic range imaging is achieved for the strong radio sources \citep{BhatnagarHDR}. The major challenge with calibration, at present, is to accurately estimate characteristics of faint and diffuse emission from a particular origin in the presence of other bright sources. We shall discuss and focus particularly on the calibration issues with the measurement of the redshifted 21-cm emission.

The temperature power spectrum of the cosmic microwave background radiation measures the statistical properties of the small fluctuations in matter density at the epoch of recombination. Following the recombination era, small fluctuations in the dark matter grew by gravitational instability. The baryonic matter, mostly neutral hydrogen roughly followed the dark matter density. Eventually, the baryonic matter density rose to the critical point and the first luminous objects were formed. This era of the cosmological evolution is termed as the cosmic dawn (CoD). Plentiful radiations from the first luminous objects ionized the neutral hydrogen, the epoch is hence termed as the epoch of reionization (EoR). In the post reionization universe, the neutral hydrogen can only be found in shelf-shielded dense compact objects, like galaxies, etc. The formation of structures, the nature of the first luminous objects and physics of reionization can be traced by studying the evolution of the brightness temperature \citep{Madau, Shaver, Furlanetto, Pritchard, Zaroubi2013}. The evolution of neutral hydrogen can be studied by observing the 21-cm line at different redshifts. Recently, \citet{BowmanEDGES21cm} reports the first observation of the cosmic dawn at a redshift of $17$. Observation of quasar absorption spectra \citep{Becker, Fan2003, Fan2006}, the optical depth for Thomson scattering from Cosmic Microwave Background (CMB) anisotropy \citep{Page, Komatsu, Hinshaw, Planck}, IGM temperature measurements \citep{Theuns, Bolton} etc suggests a redshift range of $15$ to $6$ for the EoR. 

At present several experiments are planned to estimate the power spectrum of the brightness temperature fluctuation \citep{Bharadwaj2001, Bharadwaj-EoR} from the epoch of reionization including the experiments with the GMRT \citep{Paciga}, LOFAR \citep{LOFAR}, MWA \citep{Tingay, MWA, Dillon}, the Donald C. Backer Precision Array for Probing the Epoch of Reionization (PAPER; \citet{PAPER2010, PAPER}), the Hydrogen Epoch of Reionization Array \citep{HERA}, the Square Kilometer Array \citep{SKA-Mellema, Koopmans} etc. One of the major challenges in these experiments is the presence of several orders of magnitude stronger emission at the observation frequency of the redshifted 21-cm emission (reference) \citep{Shaver, Ali-Bharadwaj-Chengalur2008, Jelic, Bernardi, Abhik}, usually termed as foregrounds. The foreground emissions include compact sources such as the radio galaxies and diffuse synchrotron and free-free emissions from the Galaxy \citep{Shaver, Matteo2002, Oh-Mack, Cooray, Ali-Bharadwaj-Chengalur2008, Abhik}. A combination of foreground avoidance \citep{Abhirup2010, TGE} and foreground mitigation \citep{Prasun} techniques have been developed.

In all the above techniques, observation of the redshifted 21-cm signal in the presence of foreground requires to measure the sky brightness distribution with a dynamic range of about $\sim 10^5$ or larger. Presence of strong foreground at the EoR frequencies additionally introduces the need for accurate calibration to achieve such high dynamic range measurements. The calibration errors often restrict the observations to reduce the expected thermal noise by integrating for a longer time. 

\citet{Abhik} subtract the foreground compact sources and estimate the power spectrum of the sky brightness distribution at 150 MHz with the GMRT. At the wavenumbers of $k \sim 0.12 - 1.2 $ h Mpc$^{-1}$, where the effect of the galactic synchrotron radiation is subdominant, they find that the power spectrum amplitude to be $\sim 1000 $ mK$^2$ with $\sim 10$ hours of observations. This is much larger than the expected redshifted 21-cm signal, they attribute it to the systematic of visibility measurements. These measurements hence pose an upper limit to the redshifted 21-cm power spectrum amplitude. \citet{Paciga} used the GMRT to observe the 150 MHz sky for 40 hours. They report an upper limit of $( 248\ {\rm mK})^2$ at wavenumbers of $0.5$ h Mpc$^{-1}$. \citet{Barry2019} have reported an upper limit of power spectral amplitude as $3900\ {\rm mK}^2$ at the wavenumber of $0.2$ h Mpc$^{-1}$ with $21$ hours of the MWA observations at the redshift $7.1$. Recently \citet{Mertens2020} gives an upper limit of $( 73\ {\rm mK})^2$ at the wavenumber of $0.075$ h Mpc$^{-1}$ at the redshift of $9.1$ with $141$ hours of observations with the LOFAR. All the above estimates of the upper limits are larger than what could be achieved with ideal calibration. This shows that at present the observations are rather limited by different systematics including uncorrected instrumental effects.

 \citet{Gehlot} have studied calibration effects such as gain errors, the effect of the polarized foregrounds, and ionospheric effects in power spectral analysis with the LOFAR-LBA. \citet{Patil} has also studied the systematic bias in the data resulting from the calibration in context to the LOFAR-EoR experiments. Other literatures in the field, such as \citep{Asad2015, Asad2016, Asad2018, Vedantham2015, Vedantham2016, Mevius}) investigate the effect of polarization leakage, ionospheric effects, etc. Effect of antenna beam variations for the observation of 21-cm signal  is studied by \citet{2020MNRAS.492.2017J}.  Effect of inaccurate models for sky-based self calibration are discussed in \citet{2016MNRAS.461.3135B,2017MNRAS.470.1849E}. An alternative to self-calibration is the redundancy calibration technique where  a priory model of the sky is not required \citep{Noordam,Weiringa}. The redundancy calibration requires existence of redundant baselines, and hence is more effective for a certain type of array design. In this calibration, the gain solutions are independent of sky models, however, the overall amplitude and phase gradients have to be set with external information \citep{Weiringa,Liu2010}. There are caveats to this method. Non-redundancy in the baseline distribution results in spectral structure that contaminates EoR detections \citep{Liu2010}.  Effect of Position errors and Beam variations on calibration solutions have been studied in literature \citep{Joseph2018,Orosz2019}. \citet{Byrne2019} show that limitations of sky based calibration results in  a fundamental limit on the calibration accuracy and  introduces additional spectral structure.

A major part of the reported upper limit of the power spectrum is due to the uncorrected gain errors in the observations. These errors appear due to the various effects, including and not restricted to the ionospheric density variation with time, variation of the instrumental gain at the small time and frequency intervals, etc. \citet{Abhirup2009, Abhirup2010} have investigated the effect of errors in the source positions arising due to residual gain in interferometric calibration. They use different but constant position errors for different sources in the sky and estimate its consequence in the systematics of 21-cm power spectrum estimation. 
In this work, we address the problem by modeling time-dependent residual gain errors directly and investigate the systematics that may arise in the observations of the redshifted 21-cm emission in presence of strong foreground, a particular antenna configuration and a gain model. We considered the gain models with time-correlated residual gains. The effect of the residual gain variation across different frequency channels and the effect of different gain models will be presented in a following paper.

The rest of the paper is arranged in the following way. In section 2 we discuss the power spectrum estimation, gain error models and their effect on the power spectrum estimation analytically. We show results from a simulated observation in section 3. The importance of the result is discussed in section 4.

\section{Analytical Calculation}
\label{sec:Analytical}
\subsection{Measurement of Visibility function}
We consider an interferometric observation, where all the antenna looks at a fixed part of the sky. Thus the antennae track the part of the sky as time progresses.
Each antenna of a radio interferometer records the electric field incident on its cross dipoles modified by the complex gain arising from the entire electronic chain. In addition to the electronic chain, the ionosphere also modifies the complex signal, and the effect is corrected often as a part of the gain calibration. The interferometers evaluate the spatial coherence function of the source or the visibility as a function of the inverse angular scale in the sky or baseline. The visibilities are calculated by cross-correlating the electric fields from each pair of antenna using digital correlators. A pair of antenna, at a given time, generates a baseline given by their positional separation projected perpendicular to the direction of observation in units of observing wavelengths. With the help of primary calibrators and self-calibration method the antenna gains are calculated as a function of time and frequencies and the visibilities are calibrated. Note that, the gain calibration can only be done to certain accuracy depending on the quality of the primary calibrators, receiver noise, ionospheric stability, etc. We shall refer to the uncalibrated part of the gain as residual gain. The residual complex gain from antenna $A$ can be modeled as 
\begin{equation}
 g_A(t, \nu) = \left [ 1+\delta_A(t, \nu) \right ]e^{i\phi_A(t, \nu)},
 \label{eq:residualgain}
\end{equation}
where $\delta_A$ is dimensionless and $\phi_A$ is measured in radian. They are much smaller than unity and randomly distributed. 
Hence, the visibility $V_{AB}^M(t, \nu)$ measured by antenna pair $A$ and $B$ in presence of the residual gain can be written in terms of the visibility function from the sky $V_{AB}^S = < E^*_A E_B>_{T_c}$ as
\begin{equation}
    V_{AB}^M(t, \nu) = <g_A^*(t, \nu) g_B(t, \nu)>_{T_c} V_{AB}^S (\nu),
    \label{eq:MeasVis}
\end{equation}
where we have assumed that the sky signal is not correlated with the gain. Furthermore, the $V_{AB}^S (\nu)$ is a time stationary signal \citep{SynthesisImaging}. Here $_{T_c}$ corresponds to the integration time to calculate the correlation of the electric fields. In case of high dynamic range observations, the visibilities $V_{AB}^S$ will have two components, the $V_{AB}^H$, the component corresponds to the strong or high signal and a component $V_{AB}^L$ corresponding to the low signal, where $\mid V_{AB}^{S, H} \mid >> \mid V_{AB}^{S, L} \mid $. In such case, it is required to have $\mid V_{AB}^R (t) \mid < V_{AB}^{L} $, where the residual visibility $V_{AB}^R$ is defined as
\begin{equation}
    V_{AB}^R (t, \nu)= V_{AB}^M(t, \nu) - V_{AB}^S(\nu).
    \label{eq:resvis}
\end{equation}

\subsection{Gain error models}
\label{sec:gainmodel}
We assume that the quantities $\delta_A(t, \nu)$ and $\phi_A(t, \nu)$, that characterizes the residual gain errors, follow Gaussian distribution with mean zero. In this work, we assume that the residual gain errors are uncorrelated in frequency and consider the effect of time correlation only. Small asymmetry in the antenna beam pattern, parallactic angle rotation and time coherence in the electron density variation in the ionosphere are expected to introduce time correlation in the residual gains. We quantify this time correlation using two-point correlation functions of $\delta_A(t, \nu)$ and $\phi_A(t, \nu)$ as 
\begin{eqnarray}
    \xi_{A}(\tau, \nu) &=& \langle \delta_A (t, \nu) \delta_A (t + \tau, \nu )\rangle,\ \ \   \sigma_{\delta}^2 = \langle \delta_A^2 \rangle \nonumber \\
    \Phi_{A}(\tau, \nu) &=& \langle \phi_A(t, \nu) \phi_A(t + \tau, \nu)\rangle,\ \ \ \sigma_{\phi}^2 = \langle \phi_A^2 \rangle,
    \label{eq:gaindef}
\end{eqnarray}
were $\xi_A (t) = 0\ \forall\ \tau \neq 0$ corresponds to the case of no time correlation. We have assumed the parameters $\sigma$ is independent of frequency, the residual gain errors from different antennae are uncorrelated and there is no correlation between the amplitude and the phase gain errors of any antenna. 
The autocorrelation function of residual gain components can be expressed alternatively in terms of the time-averaged correlations $\Delta$ and $\nabla$ defined as
\begin{eqnarray}
    \Delta(\Gamma) = \frac{1}{\Gamma} \int_0^{\Gamma} \xi_A (t) {\rm d} t \\ \nonumber
    \nabla(\Gamma) = \frac{1}{\Gamma} \int_0^{\Gamma} \Phi_A (t) {\rm d} t,
\end{eqnarray}
which are particularly useful as the ensemble average in eqn~\ref{eq:gaindef} is performed assuming the system is intermittent. The averaging time $\Gamma$ corresponds to the integration time $T_c$ used to calculate the visibilities.

\subsection{Power spectrum estimation and bias}
\label{sec:ResPow}
The statistical nature of the signal is quantified by the power spectrum of the sky brightness distribution. \citet{Bharadwaj2001} have shown that the power spectrum of the sky brightness distribution can be estimated directly by correlating the visibilities at the same baselines. In eqn~\ref{eq:MeasVis}, we have neglected the receiver noise. In most of the radio interferometric observations, the receiver noise dominates the visibility measurements. \citet{Bharadwaj-EoR} has observed that the noise is not correlated across nearby baselines, but, the sky signal is correlated. Based on this they proposed that real part of the visibility correlation at nearby baselines can be used as the power spectrum estimator 
\begin{equation}
    \mathcal{E} [ P(\vec{U}) ] = {\mathcal Re} \left [ \langle V(\vec{U})^* V(\vec{U}+\Delta \vec{U}) \rangle \right ].
    \label{eq:psest}
\end{equation}
Here, the angular brackets denote ensemble average. In general, for a wide field of view observation, the baseline vector $\vec{U}$ has three components, usually denoted as $(u, v, w)$, where $w$ is the projection along the center of the field of view of observation. It has been shown by \citet{Prasun2010} that the effect of the $w$ term depends on the baseline configuration and the functional form of the power spectrum. They show that the $w$ term introduces a modified aperture function and for a power law spectrum of slope $-2$, its effect can be neglected if the baselines chosen to estimate the power spectra are  larger than $180 \lambda$.   \citet{Ali-Bharadwaj-Chengalur2008} compare results with visibility correlation estimated by assuming   $\vec{U} = (u,v)$ between  using all the baselines and only baselines with $w<50$ for GMRT $153$ MHz observations. They find no difference in their results. In this work, we consider the baseline vector to have two components, $\vec{U} = (u,v)$. Detail discussion about the visibility correlation estimator can be seen in \citet{Bharadwaj-EoR} and its implementations in various forms can be found in \citet{Samir}. In this work, we consider the visibility correlation at the same frequency only. We shall not write the frequency dependence explicitly henceforth, if not needed particularly.

We use the visibility based power spectrum estimator that requires to perform ensemble average over many realizations of the visibility correlation at a given baseline vector. Since we always have only one realization of the observed sky, the ensemble average is done assuming spatial ergodicity. Each antenna has a finite response over the sky. This correlates the observed signal in the nearby baselines and we may consider the visibility correlations performed over a small region near a given baseline vector as different realizations. The real part of the average of these realizations is then taken as the estimate of the power spectrum at that baseline. 

The presence of residual gain errors may introduce an additional bias in the estimation of the power spectrum by visibility correlation. This bias depends explicitly on the time and the antenna pair involved in the visibility measurements. The contribution to the bias from a particular visibility correlation $V2$ can be written in terms of the residual visibilities (see eqn~\ref{eq:resvis}) as 
\begin{equation}
    V2_{ABCD}^{R}(\vec{U}, t,\vec{U} + \Delta \vec{U},t') = V_{AB}^R(\vec{U}, t)^{*} \ V_{CD}^R (\vec{U} + \Delta \vec{U}, t').
    \label{eq:viscorr}
\end{equation}
In general different combinations of antenna pairs, e. g. antenna pair AB with antenna pair CD etc. can give rise to nearby baseline vectors.
The bias in the power spectrum is average of all the $V2^{R}$ from all the possible antenna pairs giving rise to the nearby baseline vectors.  
For the power spectrum estimator discussed here, the ensemble average in eqn~(\ref{eq:psest}) is estimated over all the visibility measurements in a circular region of radius $\mid \Delta \vec{U} \mid$ at the baseline $\vec{U}$.
The visibility correlations performed to estimate the power spectrum at the baseline $\vec{U}$ will have four different types of contributions from these visibility measurements. These are correlations of baselines assumed by the same antenna at different times, correlations of baselines with one common antenna at same or different time stamps and correlations of baselines assumed by all four different antennae at the same or different time stamps. Since we do not expect any correlation in the residual gains across the different antenna, the residual gain correlation is important only when at least one antenna is common between the two baselines correlated. These cases are discussed below.
\begin{itemize}
\item{\bf Case I} Same antenna pairs correlated at different times:
Here we consider the correlation of the same antenna pairs, say AB, at different time stamps of observations to get nearby baseline vectors, i.e, 
\begin{equation} 
    V2_{ABAB}^{R}(\vec{U}, t, \vec{U} + \Delta \vec{U}, t') = V_{AB}^R(\vec{U}, t)^{*} \ V_{AB}^R (\vec{U} + \Delta \vec{U}, t').
\end{equation}
If in the region in the baseline plane over which the visibility correlation is performed to estimate the power spectrum at a given baseline, only these types of baselines are present, the ensemble averaging to estimate the power spectrum from visibility correlation is replaced with a time averag. The averaging time is directly related to the dimension of the region in baseline space over which the average is performed, the particular baseline value as well as the source declination and telescope latitude. In general, any pair of antenna traces elliptical tracks in the $u-v$ plane because of the earth's rotation. For the purpose of this discussion, we take a very simplistic model where the antenna pairs AB makes a circular arc of radius $U = \mid \vec{U}\mid$ in the baseline plane with center at $\vec{U} = 0$ and makes a complete circle in $24$ hours. The averaging time $\Gamma$ is then the time the antenna pair stays within the region of with $\Delta U = \mid \Delta \vec{U} \mid$ over which the nearest baseline averaging is done. Hence, $ \Gamma = \frac{ \Delta U \ T_{24}}{\pi U}$, where $T_{24}$ is the time in a day. Usually, the values of $\Delta U$ are kept constant for the entire baseline plane, hence, $\Gamma \propto \frac{1}{U}$. 
Assuming the gain model given in section~\ref{sec:gainmodel} and neglecting all high order correlations, the residual power spectrum for this case can be written as
\begin{equation}
    P_I^R(\vec{U}) = [\Delta_A(\Gamma) + \Delta_B(\Gamma) + \nabla_A(\Gamma) +\nabla_B(\Gamma) ]P^S(\vec{U}),
\end{equation}
where $P^S(\vec{U})$ is the power spectrum of the sky brightness distribution. Let us assume that $n_{I}(\vec{U})$ gives the fraction baseline pairs of this type available within the correlation region. Note that, $n_{I}(\vec{U})$ depends on the baseline distribution.

\item {\bf Case II} Baseline pairs with only one common antenna, correlated at the same time:
Here we consider the correlation of the baselines where one antenna is common between them, say AB and antenna AC is correlated at the same time stamps of observations to get nearby baseline vectors, i.e, 
\begin{equation}
    V2_{ABAC}^{R}(\vec{U}, t, \vec{U} + \Delta \vec{U}, t) = V_{AB}^R(\vec{U}, t)^{*} \ V_{AC}^R (\vec{U} + \Delta \vec{U}, t).
\end{equation}
Since, in this case, different time correlation between the baselines are not involved the residual power spectrum does not depend on the autocorrelation functions of the gain. The residual power spectrum can be given as
\begin{equation}
    P_{II}^R(\vec{U}) = [\sigma^2_{\delta} + \sigma^2_{\phi}]P^S(\vec{U}).
\end{equation}
Let us assume that the fraction of such baseline pairs available within the correlation region is $n_{II}(\vec{U})$.
\item {\bf Case III} Baseline pairs with only one common antenna, correlated at different times:
This is same as the case above but the correlation is done at different time stamps of observations to get nearby baseline vectors, i.e, 
\begin{equation}
    V2_{ABAC}^{R}(\vec{U}, t, \vec{U} + \Delta \vec{U}, t') = V_{AB}^R(\vec{U}, t)^{*} \ V_{AC}^R (\vec{U} + \Delta \vec{U}, t').
\end{equation}
The residual power spectrum can be written as 
\begin{equation}
    P_{III}^R(\vec{U}) = [\Delta_A(\Gamma) + \nabla_A(\Gamma) ]P^S(\vec{U}),
\end{equation}
and the fraction of such baseline pairs available within the correlation region is $n_{III}(\vec{U})$.

\item {\bf Case IV} Baseline pairs with no common antenna:
In the correlation regions, we also expect to perform the correlation between baseline pairs with no antenna in common. With our present assumptions, these visibility correlations will have zero residual power spectrum.
\end{itemize}

We assume that the autocorrelation functions and hence the $\Delta$s have power-law dependence in time, that is $ \Delta(\Gamma) = \Delta_{A_0} \Gamma^{-\alpha_{\delta_{A}}} $ and $\nabla(\Gamma) = \nabla_{A_0} \Gamma^{-\alpha_{\phi_{A}}}$. 
In general, the correlation properties of the residual gain from the different antenna are expected to be different. To investigate the effect of the gain in the residual power spectrum, simulation methods need to be used.

 In order to understand the effect of various factors in the residual power spectrum, we consider a couple of toy models here. To simplify, we assume that the correlation properties of all the antenna are same and hence the antenna indices can be dropped from the above expressions. Further, for simplicity we consider that $\sigma_{\delta} = \sigma_{\phi} = \sigma$, $\Delta_{0} = \nabla_{0}$ and $\alpha_{\delta} = \alpha_{\phi} = \alpha $. These simplifications are just to reveal the effect of baseline distribution as will be clear shortly. A more realistic situation will be discussed in later sections. Note that for $\alpha=0$, the correlation time in the residual gain error is practically infinite, whereas $\alpha=1$ corresponds to very little correlation in time. With the above simplification, the bias in the visibility correlation estimator arising due to the residual gain errors can be written as
\begin{equation}
    \mathcal{B} [P] (\vec{U}) = \left ( \left [ 4 n_{I}(\vec{U}) + 2 n_{III} (\vec{U}) \right ] \Lambda_{0} U^{\alpha} + 2 n_{II}(\vec{U}) \sigma^{2} \right ) P^S (\vec{U}),
    \label{eq:bais}
\end{equation}
where $\Lambda_{0} = \frac{\Delta_0\ \Delta U\ T_{24}}{\pi} $. Clearly, bias is multiplicative. We define the ratio of the bias to the sky power spectrum as the power spectral gain, 
\begin{equation}
\mathcal{G}(\vec{U}) = \mathcal{B}[ P ](\vec{U}) /P^S (\vec{U}),
\label{eq:psgain}
\end{equation} 
which depends on the residual gain errors and the baseline distribution and is independent of the contribution from the sky. Note that the definition of the power spectral gain is independent of the gain error and baseline distribution models.
We consider the following two simplistic models for the baseline pair distributions here. Firstly, we assume different baseline pair fractions are same, i.e,
\begin{equation}
    n_{I} = n_{II} = n_{III} =n_{IV} = \frac{1}{4}.
    \label{eq:ni_uniform}
\end{equation} 
Then the bias in the visibility correlation becomes
\begin{equation}
    \mathcal{B}[ P ](\vec{U} = \left [ \frac{3}{2} \Lambda_{0} U^{\alpha} + \frac{1}{2} \sigma^{2} \right ] P^S (\vec{U}),
    \label{eq:bias_uniform}
\end{equation}
To access the effect of $n_{i}$ in the power spectrum bias in a more realistic situation we consider 
\begin{equation}
    n_{i} (U) = \frac{ \exp \left [ - (\gamma_i U )^2 \right ]} {\sum_{i} \exp \left [ - (\gamma_i U )^2 \right ] },
\label{eq:ni_gaussian}
\end{equation}
where $N_{0}$ and $\gamma_{i}$ are constants. It is expected that at the long baselines contribution will be more from the baseline pairs of type $I$ etc., that is, $\gamma_1 < \gamma_2 < \gamma_3 < \gamma_4$. 

\begin{figure}
    \centering
    \includegraphics[width=0.5\textwidth]{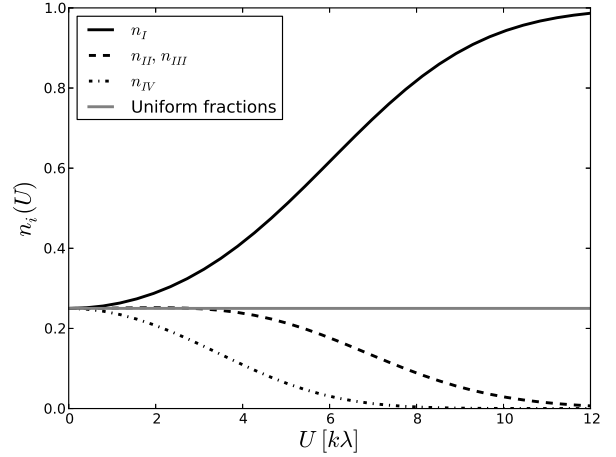}
    \caption{Models for different baseline pair distributions are shown as a function of baseline for an array with a maximum baseline of $12$ k$\lambda$. The quantities $n_{i}$ corresponds to a fraction of baseline pairs of a particular kind as discussed in section~\ref{sec:ResPow}. The horizontal grey line corresponds to the case when all types of baseline pairs have the same proportions.}
    \label{fig:fig2}
\end{figure}
Variation of the baseline pair fractions $n_{i}$ as a function of distance from the center of the baseline plane ($U$) is shown in Figure~\ref{fig:fig2}. The horizontal grey line corresponds to the case of uniform fraction given in eqn~(\ref{eq:ni_uniform}). The solid black, dashed and dot-dashed lines corresponds to $n_{I}$, $n_2,\ n_3$ and $n_4$ respectively. We have chosen $\gamma_1 = 1/6$, $\gamma_2 = \gamma_3 = 1/4$ and $\gamma_4 = 1/3$ in this plot. Clearly, at the smallest baselines, all four types have similar contributions, whereas at the largest baseline only the baseline pairs discussed in case I contribute. As can be seen, the residual gain errors have two effects in the bias in the power spectrum estimates. Firstly, for the same range of visibility correlation $\Delta U$, the time correlation in gain error affects the long baselines more than the smaller baselines. On the other hand, the fraction of baseline pairs of type I, which produces relatively more bias in the power spectrum estimates, comes in larger fraction at the long baselines. These two effects sum up and produce baseline based bias in the power spectrum estimates. 
\begin{figure}
\centering
\includegraphics[width=0.5\textwidth]{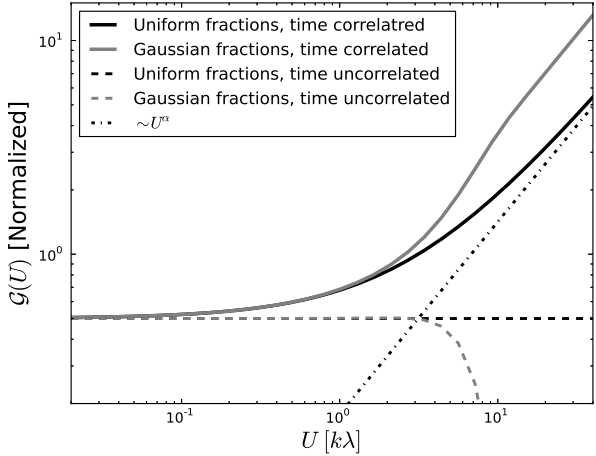}
\caption{Power spectral gain is plotted as a function of baselines (solid lines) for two types of baseline pair fractions discussed in the text. The values in the $y$ axis are normalized by the standard deviation of the gain errors. The horizontal dashed line corresponds to the case when there is no time correlation in the residual gain errors. The dot-dashed line is parallel to the autocorrelation spectra of the gain errors (see text for details). A significant change in the power spectra is expected for such time-correlated gain errors.}
\label{fig:fig3}
\end{figure}
We use eqn~\ref{eq:bais} along with the two models for the baseline pair distribution to estimate the bias in the power spectrum estimates. Figure~\ref{fig:fig3} plots the bias in the power spectrum estimates as a function of baselines. The bias is normalized by $\sigma^2 P^S(U)$. We have taken the value of $\alpha= 0.9$. The dark black line corresponds to the case with the uniform baseline pair fraction. For this case, at the short baselines, the normalized bias is almost constant at a value of $1/2$ and assumes a power-law parallel to $ U^{\alpha}$ at long baselines (dot-dashed line). The grey solid line corresponds to the case with the baseline pair fractions corresponding to eqn~\ref{eq:ni_gaussian}. Clearly, an extra steepening of the bias, due to the presence of a large fraction of baseline pair of type I at the long baselines, is seen. The dashed lines correspond to the cases when there is no time correlation in the residual gain errors. The uniform baseline pair fraction is shown in dark black and the baseline pair fractions as given by eqn~\ref{eq:ni_uniform} are shown in grey. These toy models demonstrate that the bias in the power spectrum originating from the residual gain error is scale (or baseline) dependent. The bias to signal ratio in the power spectrum estimates is directly proportional to the standard deviation of the residual gain errors and limits the dynamic range to which the estimated power spectrum has statistical significance. Clearly, if residual gain errors are smaller than unity, any signal is always larger than the bias produced for its presence. 

\section{Residual Gain Errors in the redshifted 21-cm observation in presence of extragalactic point sources}
\label{sec:Simulation}
In this section, we consider the effect of residual gain error in the measurement of the redshifted 21-cm power spectrum in the presence of strong foreground sources.
Studying the power spectrum of column density distribution of neutral hydrogen by using redshifted 21-cm radiation holds the key to probe the evolution of the structures over cosmic time and answer important questions of dark energy evolution etc. There are emissions from the galactic and extragalactic sources at the frequency ranges of the redshifted 21-cm signal. These emissions, usually several orders of magnitude higher than the 21-cm signals of interest, are usually referred as the foregrounds to the redshifted 21-cm signal. The galactic foreground is mainly a diffused synchrotron component. Though the statistical nature of this signal has been investigated in detail, the sky brightness distribution of the galactic synchrotron emission is yet to be constrained to a high dynamic range \citep{Jelic}. 

The extragalactic compact sources can be imaged using interferometric observations to a fairly high accuracy. Several techniques of modeling the visibilities from the compact sources and subtraction of their contributions from the visibilities have been discussed in literature \citep{Prasun}. In the rest of the paper, we investigate the effect of time-correlated residual gain errors in the presence of compact sources. We model the source distributions from their observed differential source count. Clustering of the compact sources is neglected here. 

The bias in the power spectrum arising due to the residual gain errors and the compact extragalactic sources are expected to be several orders of magnitude higher than the expected redshifted 21-cm signal. Since the 21-cm signal itself is expected to be quite low compared to that from the compact sources, we have neglected its own bias. 

\subsection{Point source sky model}
\label{sec:pointsrc}

The number of sources in a flux density range $S$ and $S+{\rm d}S$ per unit solid angle in the sky for a given field of view of observation is defined by the quantity $\frac{{\rm d} N}{{\rm d} S}$, known as the differential source count. The gross variation of flux density of radio sources is expressed in terms of the spectral index. The flux density of a given source as $F_1$ and $F_2$ at frequencies $\nu_1$ and $\nu_2$, the spectral index is defined as $- \frac{log(F_1/F_2}{log(\nu_1/\nu_2)}$.

The differential source count at different radio frequency bands are estimated in various surveys.
These full-sky surveys include the NRAO VLA Sky Survey (NVSS) at 1.4 GHz \citep{NVSS} Faint Images of the Radio Sky at Twenty-centimeters (The VLA FIRST Survey) \citep{FIRST} etc. At rather lower frequencies of 73.8 - 231 MHz there have been several full-sky surveys, namely the VLSS \citep{VLSS}, VLSSr \citep{Lane}, 8C \citep{Rees}, MSSS-LBA and MSSS-HBA \citep{Heald}, LoTTS \citep{Shimwell2017, Shimwell2019} etc. In a recent deep and wideband observation of the ELAIS N1 field with the uGMRT band 3 \citet{Arnab2} have estimated the differential source count at 400 MHz. It has been observed that the properties of the sources vary across the radio wavebands. The 21-cm signal from the epoch of reionization is expected to be redshifted to $\sim 100$ MHz and higher. Hence we are interested in source statistics near to this frequency band. Differential source counts at $\sim 150$ MHz have been calculated from various surveys in the literature. These include source counts from a single deep GMRT integration and a larger-area GMRT survey centered on the Bo{\"o}tes field \citep{Bootes1, Williams}, from the 7C survey \citep{McGilchrist, Hales}, from deep, small-area surveys with the LOFAR HBA system \citet{Williams, Mahony, Hardcastle}, from the MWA GLEAM survey \citep{GLEAM} as well as from deep but single-pointing MWA survey \citet{Hurley-Walker, Franzen2016, Franzen2019}, etc. \citet{Intema-TGSS} has used the entire TGSS survey to estimate the differential source count at 150 MHz based on the TGSS-ADR1 \citep{Intema-TGSS}. Because of the large sky coverage and hence large number statistics, the TGSS ADR1 provides the strict constraints on the shape of the source count distribution over the relevant flux density range. In this work we use the source counts from TGSS ADR1, where the differential source count is expressed over the flux density range of 5 mJy to 100 Jy as 
\begin{equation}
log_{10}(S^{5/2}dN/dS) = C_0 + \sum _{i=1}^{5}C_i\times[log_{10}(S)]^i.
\end{equation}
Here $C_i's$ are constants, values are given in \citet{Intema-TGSS}, [Table 6]. \citet{Intema-TGSS} have also reported the spectral index of the sources in the TGSS ADR1. Their reported spectral index distribution can be well approximated with a Gaussian distribution with mean $-0.73$ with the top and bottom 10 percent of the sources lying beyond $0.43$ and $-1.0$.
 \citet{Prabhakar} has performed an extensive investigation of the spectral index distribution of the sources in TGSS ADR1. They find for (all the sources) the mean spectral index is $-0.7$ with a dispersion of $ 0.2$. We use a Gaussian distribution of the spectral index with a mean of $-0.7$ and a standard deviation of $0.2$ to generate point source sky model at the desired frequency.

\begin{figure}
    \centering
    \includegraphics[width=0.5\textwidth]{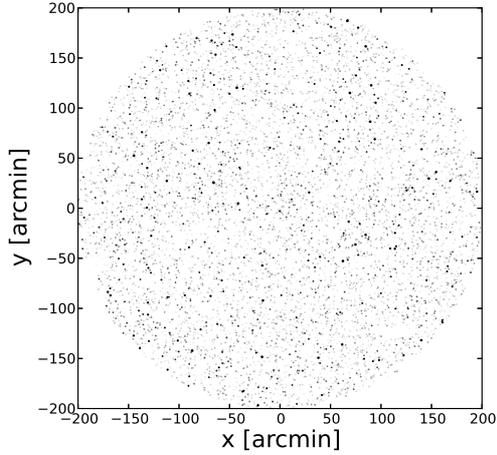}
    \caption{The point source foreground. x and y are the x positions and y positions of the sources from the pointing center in arcmin.}
    \label{fig:PSSky}
\end{figure}

\subsection{Simulation method}
\label{sec:method}
We have demonstrated the effect of baseline pair distribution on the bias in the power spectrum estimation in section~\ref{sec:ResPow} using a toy model for the telescope baseline pair distribution. 
It is expected that the residual gain errors would depend on the array configuration of a telescope. In this section, we use realistic antenna configurations to estimate the power spectrum bias. To generate the visibilities for a sky with the above source catalog, we use GMRT \citet{GMRT} baseline configuration. However, the methodology followed here can be used to do similar investigations for other arrays. Assuming a circular aperture, $45$ meter diameter GMRT dishes projects an Airy disk in the sky with the first null of this pattern at $215'$ at $130$ MHz. We use the differential source count relation and a uniform position distribution to generate a catalog of $7250$ sources in the flux density range $300\ \mu$ Jy - $1$ Jy within a circular field of view with radius $200'$. The lower limit of the flux density is comparable to the expected redshifted 21-cm signal strength at 150 MHz. As it is observed that the number of sources of flux densities over $1$ Jy is rather lower, we assume to choose a part of the sky where they are not present. 
We further assign a spectral index to each source in the catalog drawn from a Gaussian distribution with the mean and standard deviation as discussed in section~\ref{sec:pointsrc}. The distribution of the sources in this catalog is shown in Figure~\ref{fig:PSSky}.
\begin{figure}
    \centering
    \includegraphics[width=0.5\textwidth]{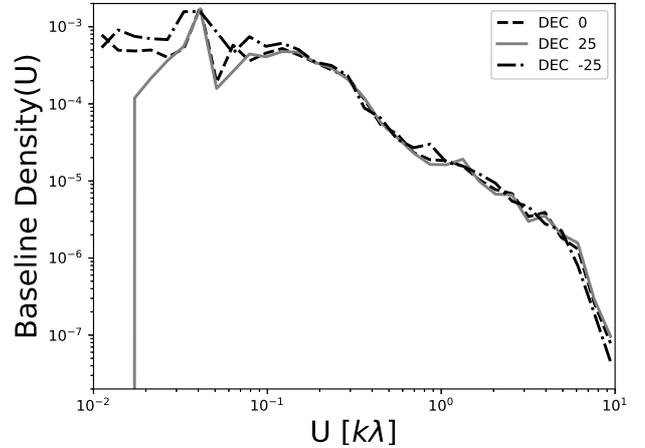}
    \caption{Variation of baseline distribution in uv-space with the different directions in the sky. Baseline density, the number of baselines per unit area at baseline U doesn't change much for different declinations.}
    \label{fig:Baseline}
\end{figure}

To compare with the expected redshifted 21-cm signal from the EoR, we choose an observation frequency of 130 MHz with a bandwidth of $8$ MHz and $128$ spectral channels. Figure~\ref{fig:Baseline} plots the azimuth averaged normalised baseline density of the GMRT at $130$ MHz for three different declination angles. The normalisation is such that the baseline distribution plotted here is independent of the total number of antennas as well as the total area in the u-v plane that the array stretch to. The baseline range of the plot reflects the available largest and smallest baselines available with the GMRT.  The smallest separation between two GMRT antenna is $\sim 100$ m, which corresponds to a baseline of $0.05\ \lambda$. Clearly, for baselines larger than $0.1$ k$\lambda$, the normalized and azimuthally averaged baseline configuration at the GMRT depends weakly on the declination angle of the source. We assume that the source distribution in the catalog discussed above is centered at the right assession of $0$ hour and declination of $25^{\circ}$ in the sky. We use the gain model as described in section~\ref{sec:gainmodel} to generate antenna based time-correlated residual gain errors. Using GMRT baseline configuration at $130$ MHz we use the VCZ \citep{van-cittert, Zernike} theorem to simulate $8$ hours of observation and generate the visibilities $V^{S}_{AB}$ for each pair of antenna $AB$ corresponding to the above sky model. We add the effect of the antenna gain to these visibilities to generate the observed visibilities $V^{M}_{AB}$ and use for further analysis. We do not add any instrument noise to the observed visibility in this study. The effect of noise will be discussed in future work.

Assuming that the sky brightness distribution of the compact sources in the sky is already modeled, we estimate the residual visibilities by subtracting $V^{S}_{AB}$ from $V^{M}_{AB}$. We use the visibility based power spectrum estimator (eqn~\ref{eq:psest}) to estimate the power spectrum from the residual visibilities for each spectral channels separately. The mean of the visibility correlation from all the spectral channels gives an estimate of the bias $\mathcal{B} [P ](\vec{U}) $ in the power spectrum in the presence of residual gain errors. Note that, in absence of any gain error, the visibility correlation ($V2$) performed with the residual visibilities have a zero mean. In the presence of residual gain errors, the visibility correlation ($V2$) from the residual visibilities measures the bias in the power spectrum.

\subsection{Results}
\label{sec:result}
\begin{figure}
    \centering
    \includegraphics[width=0.5\textwidth]{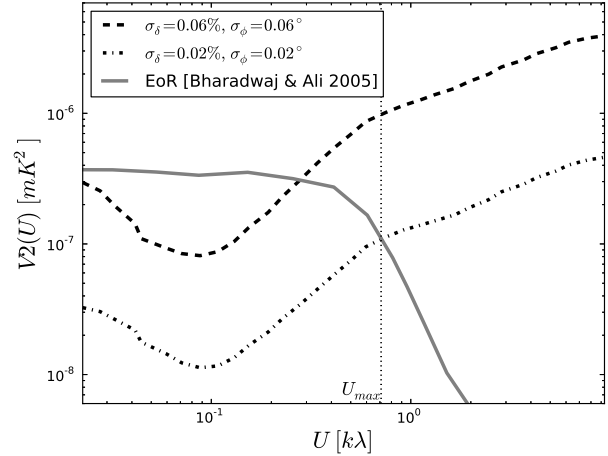}
    \caption{Residual visibility correlation plotted with baseline U. $\sigma_{\delta}$ is in percentage and $\sigma_{\phi}$ is in degree. $U_{max}$ is the point where bias in the power spectrum exceeds the EoR power spectrum, shown by the vertical dotted line.}
    \label{fig:ResPow}
\end{figure}

In our model of residual gain error, we have the parameters $\alpha_{\delta}$ and $\sigma_{\delta}$, which quantifies the amplitude of the residual gain error. The quantifiers for the phase of the gain errors are $\alpha_{\phi}$ and $\sigma_{\phi}$. The bias in the power spectrum depends on these four parameters. A value of $\alpha=0$ corresponds to the largest correlation, whereas the gain errors vanish for $\sigma=0$. We represent the strength of the amplitude gain error $\sigma_{\delta}$ and the phase gain error $\sigma_{\phi}$ in percentage and degrees respectively. Figure~\ref{fig:ResPow} shows the variation of $\mathcal{B} [P]$ as a function of $\mid \vec{U}\mid$ for $\alpha_{\delta} = \alpha_{\phi} = 0.97$. The dashed line corresponds to $\sigma_{\delta} = 0.06\%,\ \sigma_{\phi} = 0.06^{\circ}$, whereas the dot-dashed lines gives for $\sigma_{\delta} = 0.02\%,\ \sigma_{\phi} = 0.02^{\circ}$. The expected red-shifted 21-cm power spectrum (solid black line) at an observation frequency of $130$ MHz is also plotted for comparison (adopted from \citet{Bharadwaj-EoR}). Note that, the results quoted henceforth partly depend on the observation frequency chosen above. However, the analysis procedure we adopt in this work can be followed to estimate similar results at other observation frequencies. 

The bias in the power spectrum estimator follows a generic trend in Figure~\ref{fig:ResPow}, where at a baseline of $~0.1$ k$\lambda$, the bias is minimum and increases monotonically at higher baselines. The rise in the larger baselines can be attributed to a couple of facts. Firstly, the baseline coverage decreases at higher baselines and hence the chance of having more baseline pairs of type I increases, which in turn increases the bias. In addition, for the higher baselines, the time correlation in the residual gain have larger effects (see section~\ref{sec:ResPow} for detailed discussion).  We also observe that for baselines smaller than $~0.1$ k$\lambda$, the power spectrum bias increases. This happens because of the lack of shorter baselines at the GMRT. We expect this effect to be subdominant for the interferometers with better baseline coverage at shorter baselines. This generic trend in the power spectrum bias can be seen in the subsequent figures.

 Clearly, at long baselines, the power spectrum bias exceeds the power spectrum of the redshifted 21-cm emission from the HI and the residual gain errors prevent us from measuring the later. For larger values of $\sigma$ parameters, the power spectrum has a larger bias. The vertical line marks the largest baseline $U_{max}$ up to which the EoR power spectrum can be estimated in the presence of the residual gain errors for the dot-dashed curve. 
 
 At the observation frequency of $130$ MHz and at baselines $<0.5$ k$\lambda$, the redshifted 21-cm signal is almost constant (see figure~\ref{fig:ResPow}). The signal drops to half of its value at the plateau beyond $0.5$ k$\lambda$. The gain error parameters that give a $U_{max}$ value of $0.5$ k$\lambda$ or larger would allow to estimate the HI signal at these frequencies.

\begin{figure}
    \centering
    \includegraphics[width=0.5\textwidth]{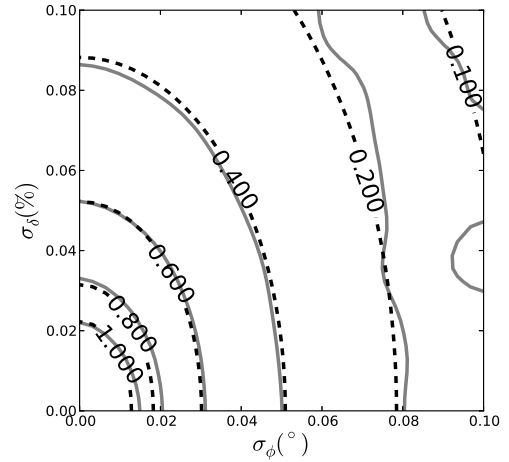}
    \caption{This contour plot is showing the variation of $U_{max}$ as a function of residual amplitude and phase gain errors. The maximum allowed baseline length $U_{max}$ in kilo wavelengths are written over contour plots. Here amplitude gain errors $\sigma_{\delta}$ and $\sigma_{\phi}$ (radian) are given in percent and degree respectively. The values of $\alpha_{\delta} = \alpha_{\phi} = 0.97$.}
    \label{fig:Umax_colour}
\end{figure}

To see the variation of $U_{max}$ with the $\sigma$ parameters we use a fiducial value of $\alpha_{\delta} = \alpha_{\phi} = 0.97$ and estimate the value of $U_{max}$ for different values of $\sigma_{\delta}$ and $\sigma_{\phi}$. Grey contours in Figure~\ref{fig:Umax_colour} gives the values of $U_{max}$ for variation of $\sigma$ parameters between $0-0.1$. Note that $U_{max}$ gives the maximum baseline to which the bias in the power spectra is lower than the power spectra itself. Hence $U_{max}$ gives the maximum baseline to which the power spectra can be estimated for each parameter values ($\sigma$, $\alpha$). A higher value of $\sigma$ corresponds to a higher residual gain error and hence is expected to provide more bias in the power spectra. It is then expected that as $\sigma$ increases, the maximum baseline to which the 21-cm signal extraction is possible, i.e $U_{max}$, decreases. This is exactly the nature we see in Figure~\ref{fig:Umax_colour}. At higher gain errors the 21-cm signal extraction is severely restricted. We have used an empirical function to fit the contours and the best fit is shown with dark dashed lines. Details of this empirical fit can be found in the Appendix I.

\begin{figure}
    \centering
    \includegraphics[width=0.5\textwidth]{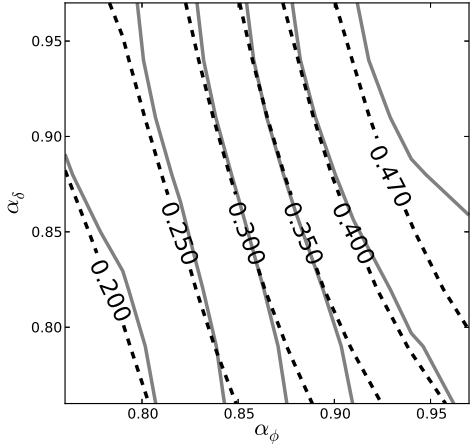}
    \caption{Contour plot showing the variation of $U_{max}$ in kilo wavelengths with the $\alpha$ parameters. The grey solid line corresponds to results from the simulation and the dark dashed lines are empirical fit to the simulation. Values of the $\sigma$ parameters fixed at $0.02$.}
    \label{fig:Umax_Alp}
\end{figure}

Figure~\ref{fig:Umax_Alp} shows the variation of the $U_{max}$ with respect to the $\alpha$ parameters for a range $0.76 - 0.97$. We have kept both the $\sigma$ parameters fixed at $0.02$. The grey contours show the values of the $U_{max}$ as calculated from the simulation. The dark dashed curves are the result of an empirical fit to the simulation results (see Appendix I). For lower values of $\alpha$ parameters, the residual gain errors are more correlated. Hence, for lower values of $\alpha$ parameters, we expect the bias in the power spectra to exceed the 21-cm signal at lower baselines. This results in a lower value of the maximum baseline, $U_{max}$, to which the 21-cm signal extraction is possible. This explains why for lower values of $\alpha$ in Figure~\ref{fig:Umax_Alp} , values of $U_{max}$ are lower.
\begin{figure}
    \centering
    \includegraphics[width=0.5\textwidth]{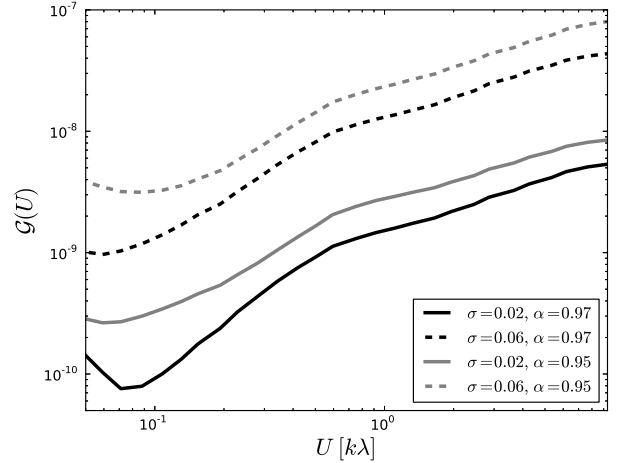}
    \caption{The ratio of the bias to the power spectrum $\mathcal{G}(U)$ for point source sky for different residual gain errors.}
    \label{fig:G}
\end{figure}

We estimate the power spectral gain $\mathcal{G}(U)$ (see eqn~\ref{eq:psgain}) for different values in our parameter space. Figure~\ref{fig:G} show the the variation of $\mathcal{G}(U)$ with baseline for $\sigma$ and $\alpha$ parameter combinations of $[0.02, 0.06]$ and $[0.95, 0.97]$ respectively. Since the quantity $\mathcal{G}(U)$ is independent of the foreground model, the estimates of the $\mathcal{G}$ can be used to calculate the bias in the power spectrum for any foreground model. 

\section{Discussion and Conclusion}
\label{sec:discussion}
In this work, we addressed the effect of residual gain errors in interferometric calibrations to estimate the power spectrum of the sky brightness fluctuations for high dynamic range observations. Our particular focus has been the estimation of the power spectrum of the redshifted 21-cm signal in the presence of strong galactic and extragalactic foregrounds. Using simulated observations with the GMRT like array configurations we found that for the nearby baseline visibility correlation estimators  the residual gain errors result in a bias in the estimates of the power spectrum. Similar bias originating from calibration errors are reported earlier in literature. The origin for the calibration errors is inaccurate sky model used in calibration \citep{Abhirup2010, 2017MNRAS.470.1849E, Patil}, bandpass calibration errors \citep{Trott&Wyth2016} etc. We discuss a methodology to access the impact of this bias given known parameters for the gain errors. We observe that the bias is scale-dependent in general. This effect adds to the systematics in the detection of the faint redshifted 21-cm signal.

The power spectrum of the sky brightness distribution is estimated by using visibility correlation at the nearby baselines \citep{Bharadwaj2001, Bharadwaj-EoR}. We first investigated analytically the origin of the bias in the power spectrum in the presence of residual gain errors through different types of baselines pairs used in the visibility correlation. We find that the bias originates when a pair of baselines are used in the visibility correlation, that has at least one antenna in common. The majority of the bias arises from the correlations done with baselines involving the same antenna pairs. Analytical calculations followed by investigations with simplistic toy models of gain errors show that the main reason for the bias is the time correlation in the residual gain errors. The bias additionally depends on the baseline distribution of the particular interferometers.

To understand the effect of the residual gain errors on the bias in the power spectrum in a more realistic scenario, we simulate visibilities for a sky with compact sources in the presence of residual gain errors. We choose a power-law function to model the time correlation in the amplitude and the phase of the residual gain errors. To distinguish the effect of the residual gain errors from the noise in the interferometer on the power spectrum bias, we do not consider any additive measurement noise here. The bias is found to increase with the standard deviation of the gain errors. It is also observed that an increase in the time correlation also increases the bias. The gain errors in the amplitude and the phase contribute approximately in a similar way to the bias.

It has been demonstrated that the compact foreground sources can be modelled accurately enough in a radio interferometric observation and can be subtracted from the observed visibilities \citep{Prasun}. However, at present, it is rather difficult to model the diffuse galactic foreground and hence a visibility based subtraction is not straight forward \citep{Jelic}. Several efforts (\citet{GLEAM} etc) are underway to generate all-sky model of the diffuse galactic foreground using radio interferometers with a rather complete baseline coverage. Such models can be used in the future for a visibility based subtraction of the diffuse foreground. In the analysis presented in this paper, we considered a point source foreground model where an exact visibility based foreground subtraction is possible. We see, however, even with a known such foreground model the estimated power spectrum will have bias coming through the residual gain errors. In our formalism, the power spectral gain $\mathcal{G}(\vec{U})$ carries the effect of the residual gain errors and is independent of the foreground model. Hence, this can be used to estimate the bias in the power spectrum in the presence of any foreground model. 

\begin{figure}
    \centering
    \includegraphics[width=0.5\textwidth]{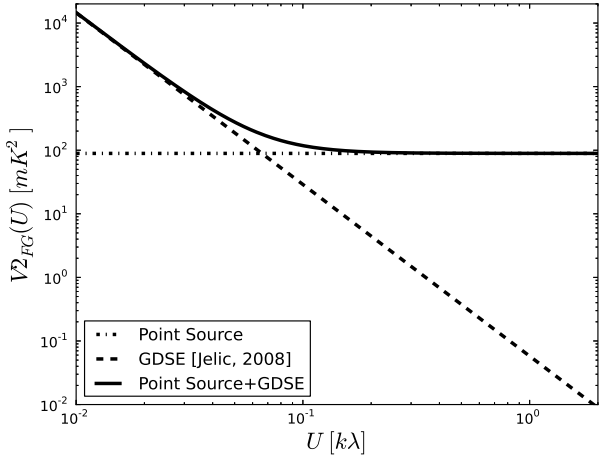}
    \caption{Power spectrum for different component of foreground. Dot-dashed line is Poisson component of point sources used in this paper, dashed line corresponds to the galactic diffuse synchrotron emission (GDSE) \citep{Jelic}. The solid curve integrates the effect of both the foreground components.}
    \label{fig:Foreground}
\end{figure}

\begin{figure}
    \centering
    \includegraphics[width=0.5\textwidth]{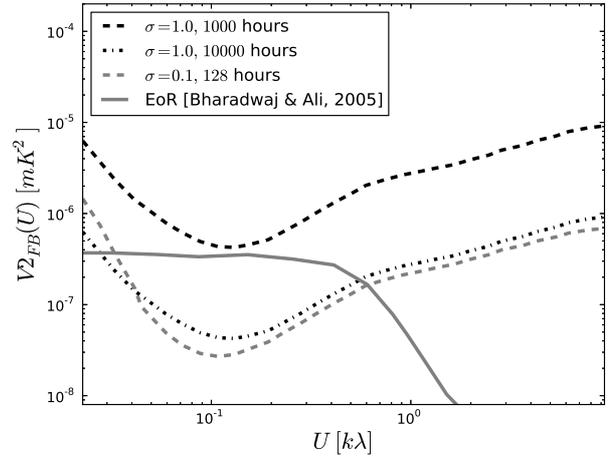}
    \caption{Bias in the power spectrum with a compact and diffuse foreground model in presence of different residual gain error models and observation time. The $\alpha$ parameters are kept at a fiducial value of $0.97$ for all gain error models. The black dashed line corresponds to $\sigma$ parameters of unity and an observation time of $1000$ hours. The black dot-dashed line corresponds to the same gain error model but with $10000$ hours of observation time. The grey-dashed line corresponds to $\sigma$ parameter set to $0.1$ and 128 hours of observation only. The grey solid line is the redshifted 21-cm signal \citep{Bharadwaj-EoR} expected at $130$ MHz.}
    \label{fig:Foreground Bias}
\end{figure}

The black solid line in Figure~\ref{fig:Foreground} shows the expected foreground power spectrum in presence of the point source model we used here (dot-dashed line) and the diffuse galactic foreground at 130 MHz \citep{Jelic} (black-dashed line). We use this full foreground model and the estimated power spectral gain $\mathcal{G}(\vec{U})$ to plot the bias in the power spectrum (black dashed line) for 1000 hours of observation with the GMRT baseline configuration in Figure~\ref{fig:Foreground Bias}. Here we have assumed that the gain errors do not have any long time correlation and they are completely uncorrelated for different days. A similar gain correlation is used by \citet{Abhirup2010}. The $\sigma$ parameters of the gain errors are set to fiducial values of $1.0$ and the $\alpha$ parameters are set to $0.97$. This value of the $\alpha$ corresponds to a low time correlation. This choice of the $\sigma$ parameter corresponds to a $1\%$ error in the gain amplitude and $1^{\circ}$ in the phase. \citet{Suman2019} have estimated that with ideal calibration and foreground subtraction the detection of the redshifted 21-cm signal at $78$ MHz with a bandwidth of $20$ MHz will require $\sim 1000$ hours of observations with the GMRT considering the available system temperatures. They also mention that a further reduction of the observation time may happen in presence of baryon-dark matter interaction \citep{Barkana, Fialkov}. \citet{Abhirup2010} uses numerical simulation to investigate the effect of residual gain error in the redshifted 21-cm power spectrum. They choose the antenna gain error amplitudes for a given antenna to be constant over one day's observation, whereas the gain errors in different days and different antenna are taken to be uncorrelated. They find that with $0.05 \%$ accuracy in the antenna gain amplitude is required to detect the redshifted 21-cm signal at a redshift of $8$ with $5000$ hours of MWA observations. In this work, we use more realistic models of residual gain errors and to estimate the bias in the power spectrum as well as the power spectral gain. The dot-dashed line in Figure~\ref{fig:Foreground Bias} corresponds to the bias in the power spectrum with the same foreground and gain model with 10000 hours of the GMRT observations and is below the expected 21-cm signal (grey solid line, adopted from \citet{Bharadwaj-EoR}) at the useable range of baselines. \citet{Mertens2020} have reported a calibration accuracy of $5\%$ for 141 hours of LOFAR observation at the redshift of 9.1. 

In this work, we see that the bias depends highly on the calibration accuracy.  Figure~\ref{fig:Foreground Bias} suggests that we may reduce the bias either by observing the signal over a larger number of days or increasing the calibration accuracy. However, now it is understood that owing to the correlated point error, mechanical beam modulation and time stability of the beam, the gain errors over different days can still be correlated and may introduce additional effects. In this view, it seems that more accurate calibration is the key to reduce the bias and detect the 21-cm signal using nearby baseline visibility correlation techniques. A reduction in the power spectrum bias is expected with only 128 hours of observations with a calibration accuracy corresponds to $0.1\%$ error in the gain amplitude and $0.1^{\circ}$ in the phase (grey dashed line in Figure~\ref{fig:Foreground Bias} ).

The standard deviation of the power spectrum estimator is also expected to change in the presence of the residual gain errors. This would add to the statistical risk in the 21-cm power spectrum estimator and need to be investigated in detail. We shall discuss the change in the variance of the power spectrum due to the presence of the residual gain errors in a companion paper.

\citet{2016MNRAS.461.3135B,Patil,Joseph2018,Orosz2019} discuss the effect of gain error in 21-cm experiments. Analytical calculations to model the gain covariance are discussed in \citet{2017MNRAS.470.1849E}. \citet{Liu2010} present a methodology to estimate the gain covariance for redundant baseline calibration.  In this work,  we assume that the gain errors are drawn from a time-correlated Gaussian distribution. The autocorrelation function of the individual antenna gain errors is assumed to follow a power law.  We use this model of the gain statistics in our methodology to estimate the power spectrum bias. We note here that for a given interferometer, careful observation has to used to establish the gain statistics first. The calibration requirements to observe the EoR signal then can be evaluated for the particular interferometer using its gain statistics. At present we are using the archival data to establish the gain statistics of the uGMRT. The results will be published as a part of a future work. As mentioned in the previous section, we use a range of $(0., 0.1)$ for the $\sigma$ and a range of $(0.76, 0.97)$ for the $\alpha$ parameters of our model of the gain statistics. These ranges for the $\sigma$ and $\alpha$ parameters is motivated by the requirement for EoR detection with our model of the gain statistics. \citet{Liu2010} estimates the calibration errors expected for a redundant baseline calibration with baseline redundancies near to a square array. Based on their calculations, with a foreground signal to noise detection of  30, a 256 element array will have $\sigma$ parameter near to $0.1$. In our model, we have neglected the time cross-correlation of the gain errors from different antennae. Note that a self-calibration based gain solution for the phase uses three antennae for the phase closure. Hence,  the phase solution of the three antennae involved in the phase closure can apparently be correlated. However, note that such phase solutions are obtained for all possible triad of antennae with a given common antenna and then averaged over. Hence, it is expected the cross-correlation of gain between two antennae is rather small compared to the self-correlation. A similar argument can be evoked for the self-calibration solutions of the amplitude gains, where the closure requires four antennae. \citet{2017MNRAS.470.1849E} use analytical calculation to show that the gain autocorrelation varies roughly as the inverse of the number of the antenna in the array, whereas the cross-correlation of gain between different antenna varies as the inverse of the square of the number of antennas. These arguments show that to the first approximation we can neglect the cross-correlation of gain between different antenna. We plan to study the cross-correlation  as a part of the ongoing study with archival the uGMRT data as stated earlier. The contribution to different baseline pair fraction in the power spectrum bias has to be recalculated if the cross-correlation of the antenna gain from the different antenna is significant.

In this work, we estimate the effect of residual gain error at different baselines for visibility correlation done at a given channel. This corresponds to the estimation of the 21-cm power spectrum as a function of the wave-number perpendicular to the line of sight and integrated over all the wave-numbers parallel to the line of sight. We have assumed that the residual gain errors are not correlated across different frequency channels. In a realistic scenario, the frequency dependence of the gain is calibrated by estimating the bandpass response of the instrument \citep{SynthesisImaging}. The estimation of the bandpass response is often done using a polynomial model, which can induce correlated residual gain errors in the bandpass response. This is expected to add to the effect of the foreground wedge when the power spectrum of the 21-cm signal is estimated in the 2D wave-number plane  \citep{Abhirup2010, Parsons2012,Morales2012,Vedantham2012}. However, since the 21-cm signal from a given redshift decorrelates faster with frequency \citep{Bharadwaj2003, Bharadwaj-EoR} than the foreground signals, the effect of frequency dependence in the residual gain errors is not straight forward to estimate. It has been observed in the literature that for the spectrally smooth gain solutions the bias in the power spectrum estimator is relatively less significant \citep{2016MNRAS.461.3135B, Trott&Wyth2016, 2017MNRAS.470.1849E}. It is still important to investigate, for the nearby baseline visibility correlation estimators how much variation in the calibration errors in frequencies affect the power spectrum bias.  We are investigating this effect at present and the results will be communicated in future work.

\section*{Appendix I}

We use the following function to model the $U(\sigma_{\delta}, \sigma_{\phi})$ as obtained
from simulation for fixed values of $(\alpha_{\delta}, \alpha_{\phi})$:
\begin{eqnarray}
  U_{max}(\sigma_{\delta}, \sigma_{\phi}) = A_0\ +& A_1\exp \left[- b \sqrt{\sigma_{\delta}^2 + \left(\frac{\sigma_{\phi}}{d}\right)^2 }\right] \nonumber \\
 +& A_2\exp \left[-c \left({\sigma_{\delta}}^2 + \left(\frac{\sigma_{\phi}}{d}\right)^2\right) \right].
\end{eqnarray}
The best fit values of the parameters and corresponding errors are given in Table ~\ref{tab:sigbf}. 
\begin{table}
	\centering
	\caption{Values of the fitting parameters defined in equation 23. Values are given in the second row and the corresponding errors in the parameter are in the third row.}
	\label{tab:sigbf}
	\begin{tabular}{l|c|c|c|c|c|c|}
\hline
& $A_0$ & $A_1$ & $A_2$ & b & c & d \\ 
& ($\times 10^{2}$) & & $(\times 10^{1})$ & & & \\
\hline
values & 6.35 & 2.20 & 6.46 & 9.35 & 9.18 & 0.577 \\ \hline
errors & 0.67 & 0.03 & 0.07 & 0.10 & 0.16 & 0.006 \\ \hline
\end{tabular}
\end{table}

We use the following function to model the $U(\alpha_{\delta}, \alpha_{\phi})$ as obtained from simulation for fixed values of $(\sigma_{\delta}, \sigma_{\phi})$ :
\begin{equation}
  U_{max}(\alpha_{\delta}, \alpha_{\phi}) = \sum_{m=0}^1\sum_{n=0}^2 a_{mn} \alpha_{\phi}^m \alpha_{\delta}^n.
\end{equation}
The best fit values of the parameters and corresponding errors (designated by $\Delta$) are
\begin{equation*}
a_{mn} = 
\begin{bmatrix}
-12.8 & 17.5 \\
 31.6 & -43.9 \\
 -19.6 & 28.1
\end{bmatrix}
\quad 
\Delta a_{mn}= 
\begin{bmatrix}
1.9 & 2.2 \\
4.2 & 4.9 \\
2.3 & 2.7
\end{bmatrix}.
\end{equation*}

\section*{ACKNOWLEDGMENT}
JK would like to acknowledge the University Grant Commission (UGC), Government of India, for financial support through a Junior Research Fellowship (UGC-JRF).  PD would like to thank  Wasim Raja for useful discussions. Authors thank the anonymous referee for suggestions that have improved the presentation of the paper significantly.

\bibliographystyle{mnras}
\bibliography{references}
\end{document}